# Enhanced opposite Imbert–Fedorov shifts of vortex beams for precise sensing of temperature and thickness

GUIYUAN ZHU, BINJIE GAO, LINHUA YE, JUNXIANG ZHANG, AND LI-GANG WANG*

*School of Physics, Zhejiang University, Hangzhou 310027, China*
*lgwang@zju.edu.cn*

**Abstract:** Imbert–Fedorov (IF) shift, which refers to a tiny transverse splitting induced by spin-orbit interaction at a reflection/refraction interface, is sensitive to the refractive index of a medium and momentum state of incident light. Most of studies have focused on the shift for an incident light beam with a spin angular momentum (SAM) i.e., circular polarization. Compared to SAM, orbital angular momentum (OAM) has infinite dimensions in theory as a new degree of freedom of light and plays an important role in light-matter coupling. We demonstrate experimentally that the relative IF shifts of vortex beams with large opposite OAMs are highly enhanced in resonant structures when light refracts through a double-prism structure (DPS), in which the thickness and temperature of the air gap are precisely sensed via the observed relative IF shifts. The thickness and temperature sensitivities increase as the absolute value of opposite OAMs increases. Our results offer a technological and practical platform for applications in sensing of thickness and temperature, ingredients of environment gas, spatial displacement, chemical substances and deformation structure.

## 1. Introduction

The specular reflection and refraction of light beam are well-known behaviors of light in geometric optics. While the non-specular reflection effects emerge from the spin-dependent excitation of electromagnetic waves at an interface, exhibiting spatial shifts named as Goos-Hänchen (GH) [1] and Imbert–Fedorov (IF) effects [2, 3].

The most interesting feature of the IF effect is the spin-related shift in transverse plane and is regarded as one kind of spin Hall effect of light induced by spin-orbit interaction between the spin angular momentum (SAM) and orbital angular momentum (OAM) of light [4]. For a paraxial propagating light beam, the SAM is $\pm 1\hbar$ determined by the polarization helicity for right- and left-hand circular polarizations, while the OAM includes extrinsic and intrinsic OAMs given by $\boldsymbol{r} \times \boldsymbol{P}$ ($\boldsymbol{r}$ and $\boldsymbol{P}$ are the transverse position and momentum of a beam) and $l\hbar$ ($l = 0, \pm 1, \pm 2, \cdots$ is the topological charge (TC) for a symmetrical left- and right-helicity vortex beam), respectively [5, 6].

The traditional IF or spin Hall effect of beam shifts is induced by the SAM of a Gaussian laser beam, and it is relatively small and was measured via weak measurements [7, 8]. This kind of tiny shift shows applications in sensing of displacement, refractive index [9-11]. For example, Zhou et al. realized the experimental observation of the spin Hall effect of light on a nanometal film via weak measurements, and their precision for the thickness of the nanometal film is 10 nm [9]. Researchers also measured the change of the refractive index by measuring the spin Hall effect via weak measurement [10, 11]. Meanwhile, the increasing of IF shift and its measurement sensitivity, e.g., using surface plasmon resonance technology [12-14] and waveguide technology [15], attract wide interests for temperature and displacement sensors, however these theoretical proposals are not yet demonstrated in experiments. Besides the

SAM effect, the intrinsic OAM of structured light beams can be manipulated via controllable values of *l* (*e.g.*, *l* = 10) [16], also allowing for enhancing the spin–orbit coupling with high angular momentum [17, 18]. The IF shift associated with OAM was first experimentally investigated using vortex beams with TE and TM polarizations [19]. Furthermore, it was theoretically revealed [20] and experimentally demonstrated [21] that the observed GH and IF shifts are linearly proportional to OAM. The studies of IF shifts in various interfaces were also concerned to show the enhanced effect of OAM on shifts [22-28].

The enhancement of IF shifts with OAM brings the possibility for its direct observation and supplies a convenient platform for enriching sensor devices. To control the TC for getting higher values of *l* and to design special resonant structures are experimentally reasonable to enhance the IF shift. Motivated by this consideration, in this work we propose to measure the relative IF shift that is the difference between the IF shifts of two vortex beams with *l* and -*l* TCs in the same polarization, giving an alternative way for greatly improving the sensitivity of the IF shift sensor. This difference of the IF shifts induced by opposite OAM beams we proposed is much larger than the difference of the same-OAM IF shifts induced by TE- and TM-polarized beams. Thus, we experimentally investigate the IF shift using a high-order vortex beam in a double-prism structure (DPS), which both lead to the enhancement of the relative IF shift. As we all know that the trajectory of geometric optics does not exist at all, here, measuring the relative IF shift of two vortex beams with *l* and -*l* TCs provides us a convenient and effective method to determine the change in the sensed structure. In our experiment, we have observed that the relative difference between the IF shifts of *l* and -*l* vortex beams is sensitive to both the air gap *d* and its temperature *T* of the DPS, showing the sensing ability for multiple parameters of the system. We have achieved the detected temperature sensitivity 33.388 μm/°C for |*l*| = 20. On the other hand, because of the existence of multiple resonance regions in the DPS, the extended sensing range becomes feasible through using different resonant regions, which can further enhance the relative IF difference by choosing the steeper changing resonance peaks. We believe that the enhanced IF shifts induced by spin-orbit interaction in this work can be applied to spatial displacement and temperature sensors.

## 2. Scheme and theory

In Fig. 1(a), it displays the schematic of the DPS that consists of two identical isosceles right-angle prisms with an air gap in the middle. The relative permittivity of the two prisms is $\varepsilon_1 = \varepsilon_3 = 2.2946$, which can be derived from the measurement of the total internal reflection. The thickness of the air gap with permittivity being $\varepsilon_2 = 1$ is initially measured at *d* = 4.16 μm. A monochromatic beam with wave vector $\vec{k}$ is incident on the DPS with an angle $\theta$ of incidence on the interface between the first prism and air gap. The critical total internal-reflection angle of light inside the prism is $\theta_c = 41.312°$, which can be measured precisely. Although the displacements of light occur at the reflection and transmission interfaces of the air gap, the actual measurements are operated in the transverse plane (vertical to the axis of propagation) outside the second prism. Figure 1(b) shows the diagram of the beam shifts experienced in symmetrical DPS. From Fig. 1(b), $Y_t$ and $Y_r$ are the IF shifts (i.e., the vertical components of the beam shifts) between the actual trajectories and the ideal optical trajectories of the transmitted and reflected light outside the prisms, respectively. $\theta_0$ and $\theta_1$ are the angles of the incident and refracted light from the outside into the first prism in Fig. 1(a), and are related via the equations: $\theta_1 = \frac{\pi}{4} - \theta$ and $\sqrt{\varepsilon_0} \sin\theta_0 = \sqrt{\varepsilon_1} \sin\theta_1$.

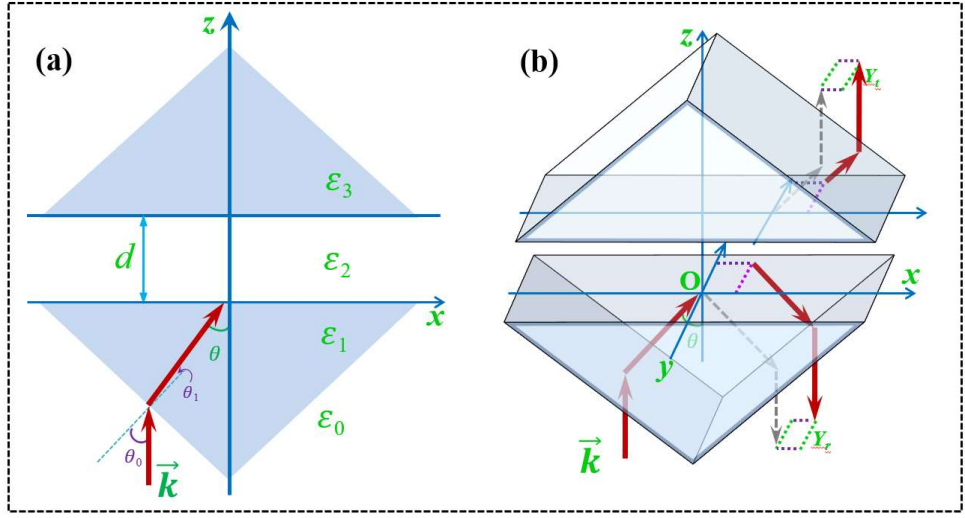

**Fig. 1.** Schematic of beam displacements in the DPS. (a) The structure of the DPS and the parameters of the incident light. Here $\vec{k}$ is the wave vector of the incident light, $\theta$ is the angle of incidence at the interface between the first prism and the air gap, $\theta_0$ and $\theta_1$ are the incident and refracted angles from the outside to the first prism, respectively, $\varepsilon_0$, $\varepsilon_1$, $\varepsilon_2$, and $\varepsilon_3$ are the relative permittivities of the outside, the first prism, the air gap, and the second prism, respectively, and $d$ is the thickness of the air gap. (b) illustrates the beam shifts in the symmetrical DPS. $Y_t$ and $Y_r$ are the IF components (vertical to the incident plane) of shifts between the actual trajectories (red) and geometrical (grey) trajectories of the transmitted and reflected light outside the DPS, respectively.

The transmission coefficient of light in this DPS is expressed as

$$t = \frac{2q_1 q_2}{q_2(q_1+q_3)\cos(k_{z2}d) - i(q_1 q_3 + q_2^2)\sin(k_{z2}d)}, \quad (1)$$

where $k_{zj} = \sqrt{k^2 \varepsilon_j - k_x^2}$ is the $z$ component of the wave vector for $k^2 \varepsilon_j > k_x^2$ in the $j$th layer with $j$ = 1, 2, 3, otherwise $k_{zj} = i\sqrt{k_x^2 - k^2 \varepsilon_j}$ for $k^2 \varepsilon_j \leq k_x^2$, $k_x = k\sqrt{\varepsilon_1}\sin\theta$ is the $x$ component of the incident wave vector inside the first prism with $k = 2\pi/\lambda$, $q_j = k_{zj}/k$ for TE polarization. Here, we only consider the TE-polarized cases, and similar results for TM-polarized cases can be analyzed and achieved in the same method. For vortex beams with TE polarization, the OAM induced IF shift can be theoretically calculated from the following expression [19-21, 29-31]

$$Y_l = l \frac{\lambda}{2\pi |t|^2}\left[\mathrm{Re}(t)\frac{d\,\mathrm{Re}(t)}{d\theta} + \mathrm{Im}(t)\frac{d\,\mathrm{Im}(t)}{d\theta}\right], \quad (2)$$

where $l$ is the TC of vortex beams. Note that the IF shift in our cases is only contributed from the OAM effect, and higher value of $l$ leads to larger IF shift. There is no contribution from the SAM effect due to the linear polarization of incident light. From Eq. (2), it is clear that the IF shifts for vortex beams with opposite TCs are totally opposite, and their difference can be enhanced in the resonant structures. Thus, we use this difference between the IF shifts for the vortex beams with + $l$ and - $l$ TCs to sense the change of the thickness and temperature.

## 3. Experimental setup, process, results and discussions

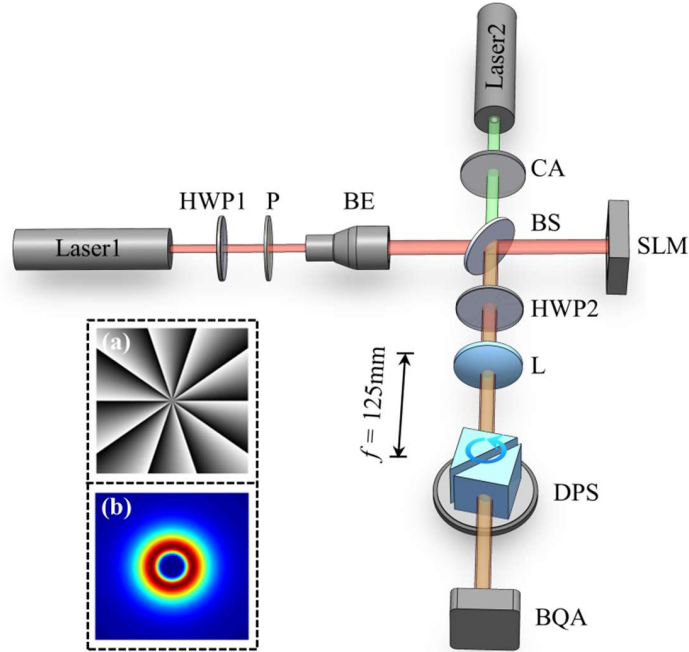

**Fig. 2.** Experimental setup for measuring IF shifts. The wavelengths of Lasers 1 and 2 are 632.8 nm and 520 nm, respectively. The notations are as follows: HWP1~HWP2, half-wave plates; P, polarizer; BE, beam expander; BS, beam splitter; SLM, spatial light modulator; CA, circular aperture; L, lens (with $f$ = 125 mm); DPS, double-prism structure placed on the electrically-controlled rotating platform; BQA, beam quality analyzer (for measuring the IF shifts). Insert (a) shows the phase diagram loaded onto the SLM and (b) the corresponding field distribution with $l$ = 10.

The experimental setup is shown in Fig. 2. A linearly-polarized laser beam 1 with wavelength of 632.8 nm passes through a half-wave plate (HWP1) and a polarizer (P). The combination of HWP1 and P is used to control the beam intensity and polarization. After the beam expander (BE), the beam diameter of the expanded light is about 3.5 mm and the expanded beam is well collimated. Then it passes through the beam splitter (BS) and is incident on the phase-only spatial light modulator (SLM) for generating different vortex beams with controllable TC $l$. The vortex beam is reflected by the SLM and then is reflected by BS. The Laser beam 2 with wavelength of 520 nm passes through the circular aperture (CA). The CA is used for removing the stray light. This beam is always adjusted to be well collimated with the propagation axis of the vortex beam (generated from the Laser beam 1) when they are coaxially propagating after the BS. We use the transmissivity curves of two light beams with different wavelengths to determine the thickness of the air gap. This double-wavelength method is more accurate than the single-wavelength method. Then the polarizations of these beams are tuned by HWP2. We use the TE-polarized beam as the incident beam to generate and measure the IF shift. Note there are similar results for the cases of TM polarization (see the Supplemental Material). The beams are focused by using the lens L (with the focal length $f$ = 125 mm) onto the interface of the air gap inside the DPS, which is placed on the controllable rotation platform to provide high precision with 1.8″ angular resolution and 36″ angle-positioning accuracy for precisely adjusting the angle of incidence. The role of this lens L here is to provide the waist position of the vortex beams at the place of the air gap of the

DPS. Thus, one can safely neglect the effect of the beam's wavefront curvature on the IF shifts due to the propagation effect of vortex beams. The transmitted beams are detected by the beam quality analyzer (BQA) (Ophir SP928, imaging CCD unit 1928×1448 pixels with each pixel size 3.69 μm × 3.69 μm), and their intensity distributions are real-timely analyzed by the commercial beam-profiling software (BeamGage). The BeamGage software is capable to determine the location of the beam center in terms of the normalized first-order moment of the measured intensity distribution [32], which gives both GH and IF shifts. Here we should point out that the distance between the DPS and the BQA is only about 8 cm, which can further avoid other effects like the displacement induced by the angular IF shifts (despite nearly zero). Here, we are only interested in IF shifts since GH shifts for different vortex beams in our cases are much smaller than IF shifts, which in turn makes the experiment become more feasible and efficient to detect and separate IF shifts. We have also found that both the change laws of IF shifts for TE and TM polarizations are similar with each other except that the amplitude for TM cases is a little smaller. Thus, in the following contents, we only discuss the results for TE polarization. Note that, the angular IF shifts in our scheme are very tiny, which theoretically induces an additional displacement about $10^{-3}$ μm (much smaller than the spatial IF shifts of vortex beams) and cancels each other thus having no contribution to the relative IF shifts we measured. Meanwhile, please see the Supplemental Material for the details of the DPS installation process, the double-wavelength method to determine thickness of the air gap, the experimental results of the IF shifts for the TE and TM polarizations, the comparison of the relative GH and IF shifts for vortex beams with opposite TCs, and the theoretically estimated magnitude of the angular IF shifts.

Since the geometrical light beams are fictitious, we have experimentally measured the relative difference between the IF shifts of the transmitted vortex beams with $l$ and $-l$. Thus, we have obtained the difference of $\Delta Y_{|l|} = Y_{|l|} - Y_{-|l|}$. In Fig. 3, it displays the IF shifts of vortex beams with $\pm l$ and their difference. In experiments, when the air gap of the DPS is fixed, we first used the vortex beams with $l=\pm 10$ to demonstrate the dependence of the IF shifts with the angle $\theta$ of incidence. The angle $\theta$ changes as the rotating platform rotates at a constant speed, meanwhile the BQA is trigged synchronously to capture every intensity distribution corresponding to every $\theta$ and it also analyzes automatically their beam-centroid positions. Then, we have measured the IF shift $Y_{10}$ for vortex beams with $l=10$ as a function of $\theta$ when $\theta$ increases from 39.45° to 41.25°, and the sampling interval of $\theta$ is 15″. In order to overcome the random fluctuations (albeit very small), we still repeated every measurement 5 times and obtained the average beam-centroid positions taken average from 5 realizations. Through the SLM, we have changed the value of $l$ and have repeated the above process. Finally, we have achieved the relationship between $\Delta Y_{|l|}$ and $\theta$, as shown in Fig. 3. Note that, in our experimental data there are the statistical fluctuation errors in the range of ±1.5 μm, which is roughly matched with the size of each pixel of the CCD.

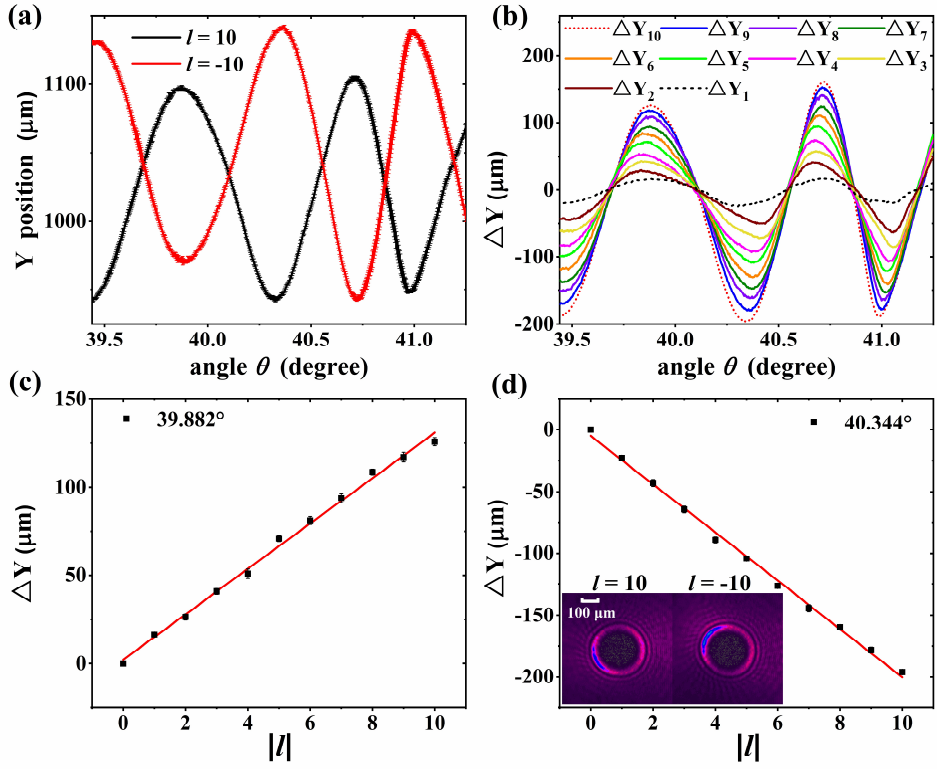

**Fig. 3.** The IF shifts of vortex beams with opposite TCs and their relative differences. (a) The IF shift ($Y_l$ position) of the vortex beams with $l = 10$ (black curve) and $l = -10$ (red curve). (b) The relative difference of IF shifts with opposite TCs (i.e., $\Delta Y_{|l|} = Y_{|l|} - Y_{-|l|}$). Here $\Delta Y_1$, $\Delta Y_2$, ..., $\Delta Y_{10}$ are denoted by different color curves. (c-d) Dependence of the relative shift $\Delta Y$ on the value of TCs at the fixed angles $\theta = 39.882°$ (c) and $\theta = 40.344°$ (d). Inset in (d) displays the intensity profiles of the transmitted vortex beams through the DPS, which are measured at the distance about 8 cm from the air gap of the DPS. The white scale bar in inset of (d) is 100 μm. The red lines in (c) and (d) are linear-fitted lines. In all cases, the thickness is fixed at $d = 4.16$ μm during the measurements.

From Fig. 3(a), it shows that both the IF shifts of opposite vortex beams are contrary due to opposite values of TCs and oscillate with the angle of incidence due to multiple resonances in the DPS. Here the thickness of the air gap in the DPS is fixed at $d = 4.16$ μm (measured from the double-wavelength method). The black and red curves, respectively, denote the values of $Y$ positions for the transmitted vortex beams with $l = 10$ and $l = -10$. It is clear that when $Y_{10}$ reaches the peak near $\theta = 39.88°$, $Y_{-10}$ nearly becomes the valley at the same angle, and as $Y_{10}$ goes to the valley close to $\theta = 40.35°$, $Y_{-10}$ trends to become the peak at almost same angle. Obviously, both $Y_{10}$ and $Y_{-10}$ curves have the opposite tendency with respect to the angle of incidence. This phenomenon caused by OAM provides us with a new way to measure IF shifts. The measured relative difference between opposite IF shifts due to opposite TCs is much larger than the relative difference between IF shifts due to different polarizations. From Fig. 3(b), it is clear seen that the relative difference of opposite IF shifts from $\Delta Y_1$ to $\Delta Y_{10}$ increases as $|l|$ increases. The maximal difference for $|l| = 10$ reaches -196 μm near $\theta$ = 40.35°. Here, we emphasize that the amplification factor is the maximum not at the angles of resonances but at the angles where the absolute slopes of the transmission curve are maximal at both sides of the resonance peaks. In Figs. 3(c) and 3(d), the relative difference is linearly proportional to the value of |l| at two different angles at both sides of a resonance peak, and the large amplification factors (the slope of the linear-fitted red lines) are achieved, showing 12.855 μm/TC and -19.535 μm/TC, respectively. More information on the amplification factor is further explained in the Supplemental Material. It turns out that using vortex beams with larger |l| is more beneficial for measuring the relative difference because of enhanced opposite IF shifts. Note that the relative difference between GH shifts for opposite vortex beams in our cases are much smaller (almost independent of the value of TCs, see the Supplemental Material) since the wide-width vortex beams are used. The inset of Fig. 3(d) also shows the transmission intensity distribution for vortex beams with l = 10 and l = -10. In the Supplemental Material, we have further pointed out that the relative difference $\Delta Y_{|l|}$ here becomes positive as the angle of incidence approaches the resonant angles and it becomes negative as the angle is away from the resonant angles. The change of IF shifts for vortex beams becomes maximal when the transmission curve changes quickly, whose underlying nature is mainly due to that the IF shifts of vortex beams here originate from the angular GH shifts [21] that are proportional to the slopes of the transmissivity or reflectivity curves.

In Fig. 4, it experimentally shows the dependence of the relative IF shifts of the vortex beams with l = ±10 on the thickness of the air gap in the DPS at different angles of incidence. Here the surrounding temperature of the DPS is tried to keep unchanged at T = 21 °C. In Fig. 4(a), when $\theta$ = 40.087°, $\Delta Y_{10}$ linearly increases with d increasing from 4.8 to 5.2 μm, and the slope of the linear-fitted line is 454.741, which is a linear amplification factor for thickness sensing. When $\theta$ = 40.150°, $\Delta Y_{10}$ increases linearly with d increasing from 4.8 to 5.4 μm, and meanwhile the slope of the linear-fitted line in this case is 383.646. When $\theta$ = 40.344°, $\Delta Y_{10}$ linearly increases with d increasing from 5.2 to 5.8 μm, and the corresponding slope of the linear-fitted line is 413.630. In Fig. 4(b), $\Delta Y_{10}$ linearly decreases with d increasing from 5.45 to 5.8 μm at $\theta$ = 40.087° and $\theta$ = 40.150°, and the slopes of the linear-fitting lines are -509.979 and -450.729, respectively. These results tell us that, $\Delta Y$ is linearly dependent on d within certain ranges of d at different $\theta$. From the slopes (i.e, the linear amplification factors), one can sense the small change of d from the large change of $\Delta Y$, and in principle, the larger the linear amplification factor, the higher the precision for achieving the change of d. Here, from the above data, we can estimate that the precision for thickness (displacement) sensing can be less than 4 nm at |l| = 10. Thus, one can always choose a suitable value of $\theta$, which can lead to a linear relation between $\Delta Y$ and d, to obtain the information of the thickness of d through measuring the value of $\Delta Y$. In practice, one can first set the region of interest on the thickness d, then choose a suitable angle of incidence to match the range of d within the linear working region, finally measuring $\Delta Y$ determines thickness d in the linear region. Unlike the method based on surface-plasmon resonance, which has usually only one measuring range near the angle of surface-plasmon excitation, here in the DPS there exist multiple resonant regions for realizing the enhanced opposite IF shifts for vortex beams with opposite TCs, leading to a wider sensing range for measuring the change of d.

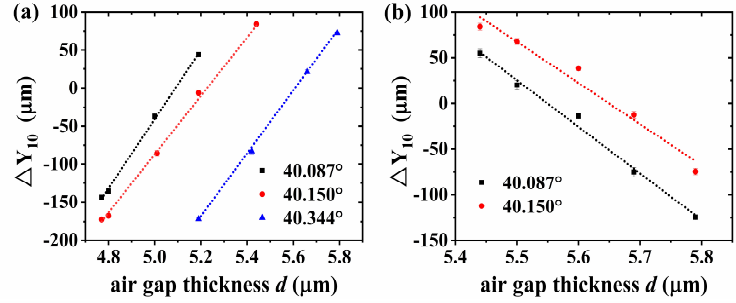

**Fig. 4.** Dependence of the relative difference $\Delta Y_{10}$ on the thickness $d$ of air gap at different incident angles in the DPS. (a) The positive and (b) negative changes of $\Delta Y_{10}$ with $d$, changing from 4.7 to 5.8 μm. Black dots are the cases of $\theta$ = 40.087°, red dots for $\theta$ = 40.150° and blue dots for $\theta$ = 40.344°. Here the surrounding temperature of the DPS is kept at $T$ = 21°C.

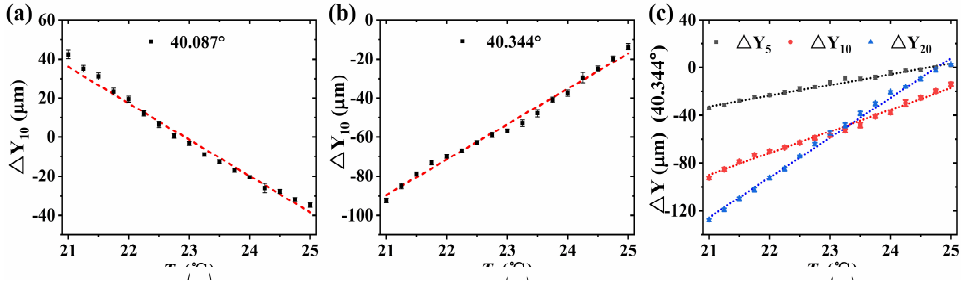

**Fig. 5.** Dependence of the relative difference $\Delta Y$ on temperature $T$ under different angles of incidence. (a) $\Delta Y_{10}$ vs $T$ at $\theta$ = 40.087°, (b) $\Delta Y_{10}$ vs $T$ at $\theta$ = 40.344°, and (c) $\Delta Y$ vs $T$ for $|l|$=5, 10, 20 at $\theta$ = 40.344°. Discrete points are measured data and dashed lines are linearly fitted. Here the thickness of air gap is initially tuned at $d$ = 5.37 μm in all situations.

Now let us turn to consider the effect of the surrounding temperature $T$ on the relative IF shifts. To investigate the effect of temperature $T$, the thickness of the air gap inside the DPS is initially adjusted to be the same value, which is confirmed by the double-wavelength method. Note that the thickness of the air gap here is initially tuned at $d$ = 5.37 μm in all situations, and a slightly increase in thickness leads to the narrowing effect of resonant peaks. Due to the change of $T$, we believe that the thickness of air gap will vary too. Since the thermo-optic coefficient for silica is $1\times10^{-5}$ K$^{-1}$ [33], the temperature dependence of prisms' refractive index is negligible. Thus, the linear change of $\Delta Y$ with temperature is mainly resulted from the effect of temperature on the thickness $d$. Here, the DPS is placed in a heating device, and a thermal sensor connected to a computer is attached to the DPS for confirming the surrounding temperature of the DPS. The precision of the thermal sensor is 0.125 °C. The temperature is displayed and recorded in real time on the computer.

Figures 5(a) and 5(b) show that $\Delta Y$ is linearly dependent on temperature at two different angles. In Fig. 5(a), when $\theta$ = 40.087° (at the right side of one resonant peak), $\Delta Y_{10}$ linearly decreases with $T$ within the range from 21°C to 25 °C, and the slope of the linear-fitted line is -18.608 μm/°C. In Fig. 5(b), when $\theta$ = 40.344° (at the left side of another resonant peak),

$\Delta Y_{10}$ linearly increases with $T$ within the same temperature range, and now the slope of the linear-fitted line is 18.115 μm/°C. In Fig. 5(c), when we increase the TCs of vortex beams, in the same temperature range and angle ($\theta$ = 40.344°), the relative differences for $\Delta Y_5$, $\Delta Y_{10}$, and $\Delta Y_{20}$ increase linearly, and their slopes increase as the value of $|l|$ increases. Here the slope for $\Delta Y_5$ is 8.915 μm/°C and it becomes 33.388 μm/°C for $\Delta Y_{20}$, so that the precision for temperature sensing is less than 0.045°C at $|l|$ = 20 in this structure. Such linear dependence shows us that using the differences between opposite IF shifts of vortex beams in the DPS can be a possible tool as a temperature sensor in microstructures.

Here we use the above slope, which can also be defined by $S = \frac{\partial \Delta Y}{\partial T}$, to estimate the sensitivity of the temperature sensor. From Fig. 6, it shows the effect of OAM on the sensitivity $S$ of the temperature sensor at $\theta$ = 40.344°. Here the thickness is initially tuned at $d$ = 5.68 μm. As the value of $|l|$ increases, the slope of $\Delta Y$ increases, thus the sensitivity $S$ increases too. Therefore, one can improve the sensitivity $S$ of this temperature sensor by increasing $|l|$. In our experiment, we utilized vortex beams with the maximal value of $|l|$=20, which corresponds to the maximum $S$ be 33.388 μm/°C as pointed out above. Of course, in practice, the increasing of the sensitivity $S$ is limited by the quality of generated vortex beams. We use the SLM to generate the vortex beams with different TCs. The quality of vortex beams will decrease as the value of TCs increases due to the limited modulation ability of the SLM. It should be mentioned that the fluctuation of the measured data in Figs. 5 and 6 is mainly due to the uncontrollable temperature fluctuations, such as microscale air-flow disturbance and heat conduction in the experimental environment.

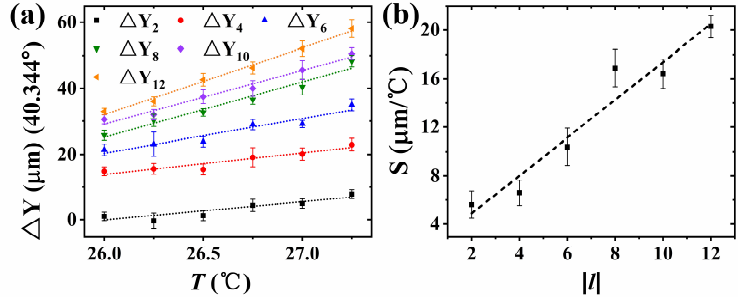

**Fig. 6.** (a) Effect of temperature on the relative difference of opposite IF shifts and (b) their temperature sensitivity $S$ under different TCs. In (a), the black, red, blue, green, purple, orange dots denote $\Delta Y_2$, $\Delta Y_4$, $\Delta Y_6$, $\Delta Y_8$, $\Delta Y_{10}$, and $\Delta Y_{12}$, respectively. Here $d$ = 5.68 μm and $\theta$ = 40.344°.

Finally, we have noted that there are some theoretical studies on the application of the IF shifts in sensing [12-17]. For example, in Ref. [13], without OAM, the temperature sensitivities of 0.79 cm/K and 188 μm/K were theoretically predicted with the GH shift and IF shift, respectively. In Ref. [14], the sensitivity of the IF shifts was theoretically found to be enhanced by increasing the incident vortex charge, and the maximal sensitivities for the GH and IF shifts can theoretically reach $3.01 \times 10^3$ μm/RIU, $4.63 \times 10^3$ μm/RIU with $l$ = 1. Yet all these theoretical predictions have not been demonstrated in experiments. Here we proposed and experimentally realized to use the relative IF shifts of vortex beams with opposite TCs for the thickness and temperature sensing in the DPS, and such method could be naturally extended to other resonant structures like photonic crystals. We believe that our work paves the route to use the IF shifts in sensing devices.

## 4. Conclusion

In summary, we have experimentally detected the relative IF shifts of vortex beams with tunable but opposite OAMs in the DPS. Our experiments have demonstrated that the relative difference between these opposite IF shifts linearly increases with the increase of TC. This method has the better advantage than that based on the difference between the IF shifts of the same TC vortex beams with different polarizations. One can use the enhanced opposite IF shifts in the DPS to realize the measurements on the change of thickness of the air gap and to achieve temperature sensing from the relative IF shifts by choosing the suitable angle of incidence. For the displacement sensor, at different incident angles, there are different linear ranges and optimal amplification factors. The displacement sensor in this work has a wide range for measuring $d$. For the temperature sensor, the sensitivity $S$ can be enhanced by using larger topological charge $|l|$, and we have measured the maximal temperature sensitivity $S$ to be 33.388 μm/°C by using the vortex beams with $|l|$=20. These results can promote the application of controlling IF shifts of vortex beams with opposite TCs in optical sensors.

**Acknowledgments:** National Natural Science Foundation of China (NSFC) (grants No.11974309 and 62375241), and National Key Research and Development Program of China (2021YFC2202700).

**References**

1. F. Goos and H. Hänchen, "Ein neuer und fundamentaler versuch zur totalreflexion," Ann. Phys. (Berlin) **436**, 333-346 (1947).
2. C. Imbert, "Calculation and experimental proof of the transverse shift induced by total internal reflection of a circularly polarized light beam," Phys. Rev. D **5**(4), 787-796 (1972).
3. F. I. Fedorov, "K teorii polnogo otrazheniya," Dokl. Akad. Nauk SSSR **105**, 465-468 (1955).
4. M. Onoda, S. Murakami, and N. Nagaosa, "Hall effect of light," Phys. Rev. Lett. **93**, 083901 (2004).
5. L. Allen, S. M. Barnett, and M. J. Padgett, *Optical Angular Momentum* (1st ed.), CRC Press. (2003).
6. V. G. Fedoseyev, "Transformation of the orbital angular momentum at the reflection and transmission of a light beam on a plane interface," J. Phys. A: Math. Theor. **41**, 505202 (2008).
7. G. Jayaswal, G. Mistura, and M. Merano, "Observation of the Imbert–Fedorov effect via weak value amplification," Opt. Lett. **39**(8), 2266–2269 (2014).
8. O. Hosten and P. Kwiat, "Observation of the spin Hall effect of light via weak measurements," Science **319**, 787–790 (2008).
9. X. Zhou, Z. Xiao, H. Luo, and S.Wen, "Experimental observation of the spin Hall effect of light on a nanometal film via weak measurements," Phys. Rev. A **85**(4), 043809 (2012).
10. J. Liu, K. Zeng, W. Xu, S. Chen, H. Luo, and S. Wen, "Ultrasensitive detection of ion concentration based on photonic spin Hall effect," Appl. Phys. Lett. **115**(25), 251102 (2019).
11. R. Wang, J. Zhou, K. Zeng, S. Chen, X. Ling, W. Shu, H. Luo, and S. Wen, "Ultrasensitive and real-time detection of chemical reaction rate based on the photonic spin Hall effect," APL Photon. **5**(1), 016105 (2020).
12. L. Salasnich, "Enhancement of four reflection shifts by a three-layer surface-plasmon resonance," Phys. Rev. A **86**, 055801 (2012).
13. Y. Xu, L. Wu, and L. K. Ang, "Ultrasensitive optical temperature transducers based on surface plasmon resonance enhanced composited Goos–Hänchen and Imbert–Fedorov shifts," IEEE J. Sel. Topics Quantum Electron. **27**(6), 4601508 (2021).
14. X. Guo, X. Liu, W. Zhu, M. Gao, W. Long, J. Yu, H. Zheng, H. Guan, Y. Luo, H. Lu, J. Zhang, and Z. Chen, "Surface plasmon resonance enhanced Goos–Hänchen and Imbert–Fedorov shifts of Laguerre–Gaussian beams," Opt. Commun. **445**, 5–9 (2019).
15. T. Tang, C. Li, L. Luo, Y. Zhang, and Q. Yuan, "Thermo-optic Imbert–Fedorov effect in a prism-waveguide coupling system with silicon-on-insulator," Opt. Commun. **370**, 49–54 (2016).
16. D. Naidoo, F. S. Roux, A. Dudley, I. Litvin, B. Piccirillo, L. Marrucci, and A. Forbes, "Controlled generation of higher-order Poincaré sphere beams from a laser," Nat. Photon. **10**, 327–332 (2016).
17. A. M. Yao and M. J. Padgett, "Orbital angular momentum: origins, behavior and applications," Adv. Opt. Photon. **3**, 161–204 (2011).
18. R. C. Devlin, A. Ambrosio, N. A. Rubin, J. B. Mueller, and F. Capasso, "Arbitrary spin-to-orbital angular momentum conversion of light," Science **358**, 896–901 (2017).
19. R. Dasgupta and P. K. Gupta, "Experimental observation of spin-independent transverse shift of the centre of gravity of a reflected Laguerre–Gaussian light beam," Opt. Commun. **257**, 91–96 (2006).

20. K. Y. Bliokh, I. V. Shadrivov, and Y. S. Kivshar, "Goos–Hänchen and Imbert–Fedorov shifts of polarized vortex beams," Opt. Lett. **34**(3), 389-391 (2009).
21. M. Merano, N. Hermosa, and J. P. Woerdman, "How orbital angular momentum affects beam shifts in optical reflection," Phys. Rev. A **82**, 023817 (2010).
22. N. Hermosa, M. Merano, A. Aiello, and J.P. Woerdman, "Orbital angular momentum induced beam shifts," Proc. of SPIE **7950**, 79500F-2 (2011).
23. W. L. offler, N. Hermosa, A. Aiello, and J. P. Woerdman, "Total internal reflection of orbital angular momentum beams," J. Opt. **15**, 014012 (2013).
24. C. Prajapati, "Numerical calculation of beam shifts for higher-order Laguerre-Gaussian beams upon transmission," Opt. Commun. **389**, 290-296 (2017).
25. W. Zhu, H. Guan, H. Lu, J. Tang, Z. Li, J. Yu, and Z. Chen, "Orbital angular momentum sidebands of vortex beams transmitted through a thin metamaterial slab," Opt. Express **26**(13), 17378-17387 (2018).
26. M. T. Lusk, M. Siemens, and G. F. Quinteiro, "Large centroid shifts of vortex beams reflected from multi-layers," J. Opt. **21**, 015601 (2019).
27. Z. Cui, Y. Hui, W. Ma, W. Zhao, and Y. Han, "Dynamical characteristics of Laguerre–Gaussian vortex beams upon reflection and refraction," J. Opt. Soc. Am. B **37**(12), 3730-3740 (2020).
28. S. Ahmad, M. Abbas, M. Awais, A. Khan, and Z. Uddin, "Influence of orbital angular momentum of vortex light on lateral shift behavior," J. Opt. **23**, 115402 (2021).
29. L-G. Wang, H. Chen, and S-Y. Zhu, "Large negative Goos-Hänchen shift from a weakly absorbing dielectric slab," Opt. Lett. **30**(21), 2936-2938 (2005).
30. A. Aiello, M. Merano, and J. P. Woerdman, "Duality between spatial and angular shift in optical reflection," Phys. Rev. A **80**, 061801 (2009).
31. K. Y. Bliokh and A. Aiello, "Goos–Hänchen and Imbert–Fedorov beam shifts: an overview," J. Opt. **15**, 014001 (2013).
32. Ophir Optronics, "BeamGage user guide," (2023), https://www.ophiropt.com/laser--measurement/beam-profilers/services/manuals.
33. J. Qin, L. Deng, J. Xie, T. Tang, and L. Bi, "Highly sensitive sensors based on magneto-optical surface plasmon resonance in Ag/CeYIG heterostructures," AIP Adv. **5**, 017118 (2015).

# SUPPLEMENTAL MATERIAL
*of*

## ENHANCED OPPOSITE IMBERT–FEDOROV SHIFTS OF VORTEX BEAMS FOR PRECISE SENSING OF TEMPERATURE AND THICKNESS

*By Guiyuan Zhu, Binjie Gao, Linhua Ye, Junxiang Zhang, and Li-Gang Wang*[*]
*School of Physics, Zhejiang University, Hangzhou 310027, China*
*\*lgwang@zju.edu.cn*

## 1. The DPS installation

The process of fabricating the air gap in our DPS is as follows. Two thin polyimide films with each thickness of 6 μm are sandwiched tightly between the left and right edges of these two prisms, creating an air gap between them under considerably large pressure by using the fastened screws. The DPS is firmly fixed to the rotation platform. A slight change in the rotation angle of the screw will affect the thickness of the air gap, so we can change the thickness of the air gap by fine-tuning the rotation angle of the screw. In our experiment, the thickness $d$ of the air gap is measured by fitting the measured transmission curve (as a function of angle of incidence) of two different wavelengths of laser beams. Here, $d$ is always smaller than the thickness of polyimide films due to the pressure from the fastened screws.

## 2. Determination of the air gap thickness by the double-wavelength method

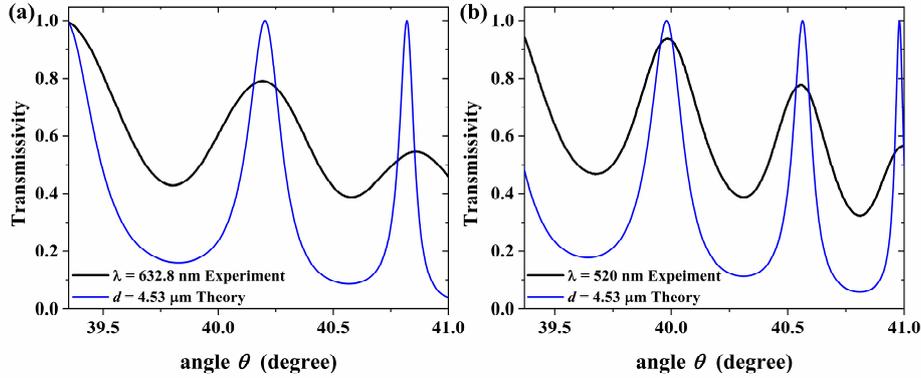

**Fig. S1.** Transmissivity of two beams with (a) $\lambda_1$ = 632.8 nm and (b) $\lambda_2$ = 520 nm. The black lines are the experimental results, and the blue lines are the theoretical simulations.

Figure S1 shows the transmissivity of two beams with $\lambda_1$=632.8 nm and $\lambda_2$=520 nm. It is seen that the two experimental results are in good agreement with the theoretical simulation. This is the double-wavelength method used to determine the thickness of the air gap. The thickness has a great influence on the angular interval of the transmittance peaks. The greater the thickness, the smaller the angular interval between the peaks. Therefore, $d$ is determined by optimally fitting the peak interval of experimental transmissivity data.

## 3. The theoretical simulation of the IF shifts of vortex beams

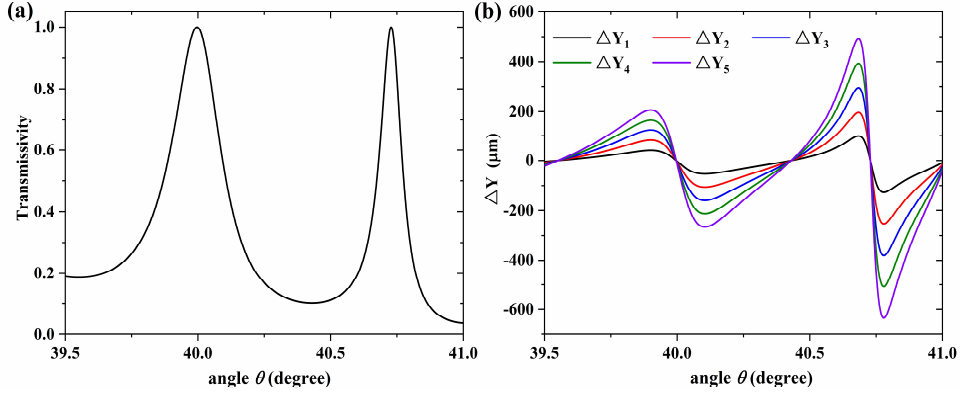

**Fig. S2.** The theoretical simulation of the transmissivity and relative IF shifts $\Delta Y$. (a) The transmissivity, (b) the black, red, blue, green, and purple curves denote the theoretical values of $\Delta Y_1$, $\Delta Y_2$, $\Delta Y_3$, $\Delta Y_4$, and $\Delta Y_5$, respectively. Here we take $d = 4.16$ μm.

According to Eqs. (1-2), we can get the theoretical results of IF shifts. Figure S2(a) shows the transmissivity of the DPS. Figure S2(b) shows the theoretical simulation of $\Delta Y_{|l|}$ with $l = 1$, 2, ..., 5. The black, red, blue, green, and purple curves denote the theoretical relative shifts $\Delta Y_1$, $\Delta Y_2$, $\Delta Y_3$, $\Delta Y_4$, and $\Delta Y_5$, respectively. We can see that, as $|l|$ of opposite TCs increases from 1 to 5, the relative IF shift $\Delta Y_{|l|}$ increases. Thus, from the theoretical point of view, the OAM does have the effect on the IF shift. Meanwhile, from Fig. S2, we can also find that the relative difference $\Delta Y_{|l|}$ here becomes positive as the angle of incidence approaches the resonant angles and it becomes negative as the angle is away from the resonant angles.

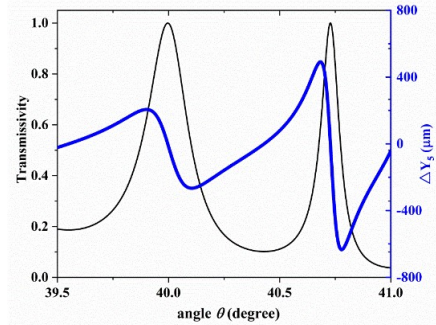

**Fig. S3.** The theoretical predictions of the transmission curve (the thin black curve, left axis) and the relative IF shift $\Delta Y_5$ (the thick blue curve, right axis). Here $\Delta Y_5$ is for the relative IF shift of vortex beams with ±5 TCs.

For the better explanations, we put the transmission curve and one of the relative IF shift together in Fig. S3. Now we can clearly find the relative IF shift becomes positive as the angle of incidence approaches the resonant angles (i.e., the value of $\theta$ is located at the left side of the resonant peak), and it becomes negative as the angle $\theta$ is away from the resonant angles (i.e., the value of $\theta$ is located at the right side of the resonant peak). According to Ref. [1] or Eq. (2), one can readily obtain the relative IF shift of vortex beams with opposite TCs as $\Delta Y_{|l|} \propto 2l\Theta_{GH}$, where $\Theta_{GH}$ is the angular GH shift of a fundamental Gaussian beam. As we know that for transmitted or reflected light, the angular GH shifts are proportional to the slopes of the transmissivity or reflectivity curves. Thus, when the transmission curve increases quickly with the increasing angle of incidence, the relative IF shift becomes the positive maximum, in contrast when the transmission curve decreases quickly with the increasing angle of incidence, the relative IF shift becomes the negative minimum. Thus, the change of the relative IF shifts between vortex beams with opposite TCs becomes maximal when the curve of transmissivity dramatically changes, and the relative IF shifts of vortex beams are zero when the angles of incidence are located at the resonant peaks or the valleys of the transmission curve. Thus, the change of IF shifts for vortex beams becomes maximal when the transmission curve changes quickly, whose underlying nature is mainly due to that the IF shifts here originate from the angular GH shifts [1].

## 4. The negligible angular IF shift for vortex beams with different TCs

According to the work in Ref. [1], the angular IF shift is given by $\Theta_{IF}^l = (1+2|l|)\Theta_{IF}$, where $\Theta_{IF}$ is the angular IF shift of a fundamental Gaussian beam. In our case, this angular IF shift for TE polarization is given by $\Theta_{IF} = -\frac{\theta_0^2}{2} \text{Re}\left[2i \cot\theta \left(\frac{t_{TM}+t_{TE}}{t_{TE}}\right)\right]$ $= \frac{-\lambda^2}{2\pi w_0^2} \text{Re}\left[2i \cot\theta \left(\frac{t_{TM}+t_{TE}}{t_{TE}}\right)\right]$, where $\theta_0$ and $w_0$ are the angular spread and the waist of the incident Gaussian beam. This quantity $\Theta_{IF}$ is very tiny and it is less than $10^{-7}$ rad in our cases, which induces an additional displacement about $10^{-3}$ μm (much smaller than the spatial IF shifts of vortex beams). Here we provide the theoretical spatial and angular IF shifts of a Gaussian beam with the TE polarization in the below Fig. S4. From Fig. S4, we can see that the value of the angular IF shift is very tiny (see Fig. S4(b)), and the spatial IF shift of the Gaussian beam is also much small (less than 1μm), which is also much smaller than the spatial IF shifts of vortex beams with $l=1$, 2, and 3 (for examples, see Fig. S4(a)).

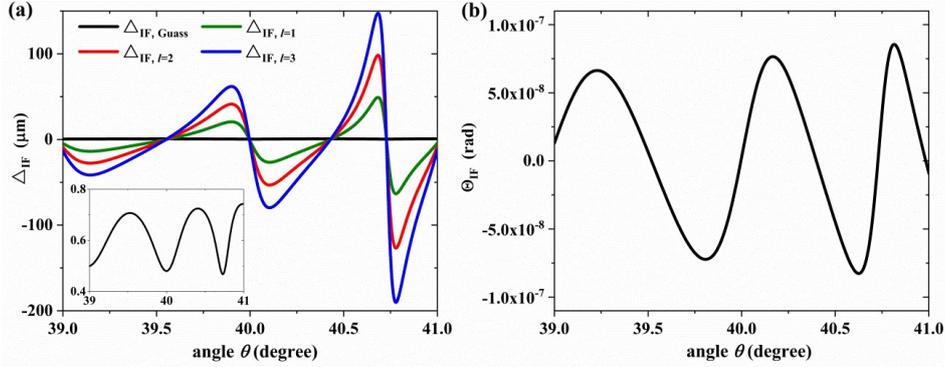

**Fig. S4.** (a) Comparison of the spatial IF shifts between a Gaussian beam (black) and vortex beams with $l = 1$ (green), 2 (red) and 3 (blue), and (b) the angular IF shift of a Gaussian beam in our double-prism system with the air thickness $d = 4.16$ μm. Here the beam parameter is $w_o = 1$ mm and the light beam is the TE polarization.

Furthermore, according to our proposal, we use the difference between the IF shifts of the vortex beams with opposite TCs as a probe of thickness and temperature sensing. Even if the displacements induced by the angular IF shifts of vortex beams with $\pm l$ TCs exist in our original data, they can be canceled since their displacements are the same in principle (only depending on the absolute value of TCs) as shown above.

## 5. Discussions of the theoretical and experimental amplification factors

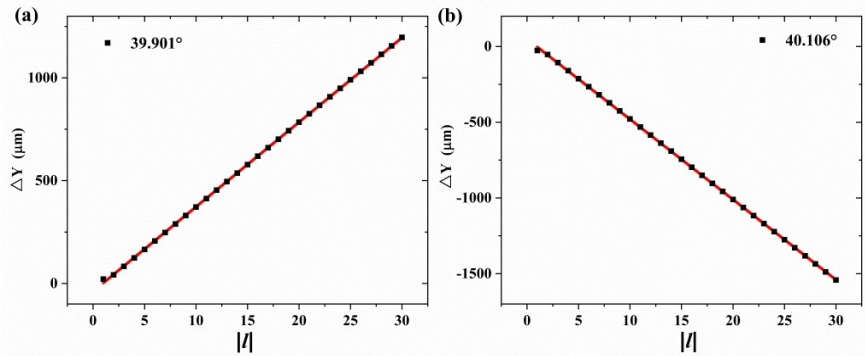

**Fig. S5.** Theoretical linear amplification relationship between $\Delta Y$ and $l$. Their amplification factors (i.e., the slopes of the linear-fitted red lines) are (a) 41.137 μm/TC at $\theta = 39.901°$, and (b) -52.991 μm/TC at $\theta = 40.106°$, respectively.

According to Fig. S2 (b), we can obtain the theoretical amplification factors at different angles. Figure S5 shows the theoretical amplification factors (that is the slope of the linear-fitted lines) are 41.137 μm/TC for the maximum $\Delta Y$ condition at $\theta = 39.901°$, and -52.991 μm/TC for the minimum $\Delta Y$ condition at $\theta = 40.106°$, respectively. Note that this amplification factor is strongly dependent on the angles in the resonant structures since the relative IF shifts are oscillating with the angles of incidence as shown in Fig. S2(b). In principle, if the angle of incidence corresponds to the peak or valley of the transmission curve, i.e., when the slopes of the transmission curve vs the angle are zero, in these situations the amplification factors will be zero. When the angle of incidence corresponds to the dramatical change of the transmission curve, theoretically the maximal value of $\Delta Y$ increases linearly as $l$ increases. Thus, the sensitivity $S$ of sensor can be enhanced by using larger topological charge.

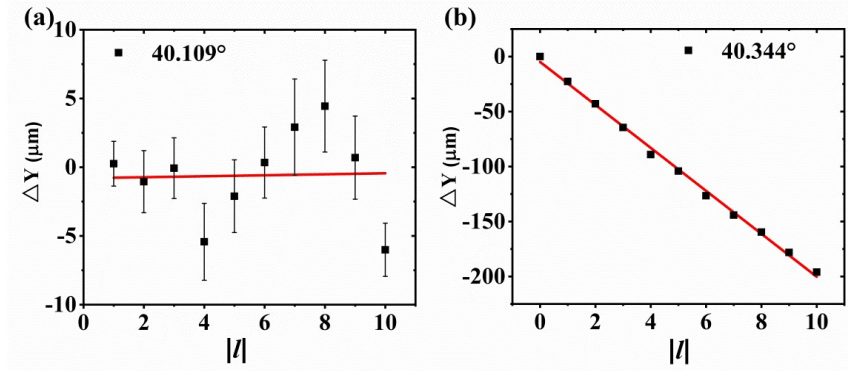

**Fig. S6**. Comparison of experimental amplification factors at two different angles: (a) the angle of practical resonance and (b) the angle where the transmission curve decreases quickly. The amplification factors (i.e., the slopes of the linear-fitted red lines) are 0.036μm/TC for the angle at $\theta = 40.109°$ (which is close to the resonant condition), and -19.535μm/TC for the angle at $\theta = 40.344°$ (where transmission curve drops quickly), respectively.

For the readers' reference, we provide the experimental data for the amplification factors in two limited cases. From Fig. S6, we can see that, near the angle of resonances the amplification factor (the slope of the linear-fitted red line) is almost zero (here it is 0.036 μm/TC), and the data are fluctuated around zero due to the experimental factors such temperature fluctuation and positioning accuracy of the angle scanning. When the angle is close to the situation that the transmission curve drops quickly, the amplification factor is -19.535 μm/TC in the case of Fig. S6(b). In our manuscript, we have provided the two situations, corresponding the quick increasing and decreasing of both sides of a transmission peak, respectively, see Figs. 3(c) and 3(d). Note that there exist differences between the theoretical predictions and experimental measurements on the angles since any slightly change in temperature will lead to the thickness fluctuation of the air gap in the DPS, inducing the shifts of the practical resonant angles.

## 6. The experimental results of the IF shifts for TM and TE polarizations

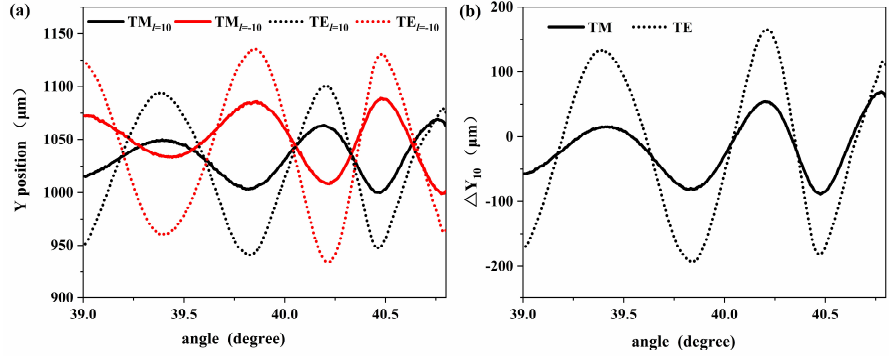

**Fig. S7.** The experimental results of Y position of the transmitted beam and relative IF shifts $\Delta Y$ for TM and TE polarizations. (a) Y position of the transmitted beam for $l = 10$ and $l = -10$ for TE and TM polarizations, (b) the relative IF shifts $\Delta Y_{10}$, the black and red curves denote $l = 10$ and $l = -10$, respectively. The solid and dotted curves denote TM and TE polarizations, respectively, and here $d = 4.55$ μm.

Figure S7(a) shows the Y position of the transmitted beam for $l = 10$ and $l = -10$, and Fig. S7(b) shows their relative IF shifts $\Delta Y_{10}$, the black and red curves denote $l = 10$ and $l = -10$, respectively, and the solid and dotted curves denote TM and TE polarization, respectively. We can find that the change law for the IF shifts of vortex beams with opposite TCs for TM polarization is consistent with those for TE polarization, but the amplitude for TM cases was smaller. Thus, in our work we use the vortex beams with TE polarization. Note that the thickness $d$ here is different from that in the manuscript and was confirmed from the above double-wavelength method. Before doing the experiment, we first confirmed the thickness of air gap in the DPS when the experiment was done in different days.

## 7. Comparison of the experimental results of the relative GH and IF shifts for vortex beams with opposite TCs

Figure S8 shows the comparison of the experimental data for the relative GH and IF shifts for vortex beams with opposite TCs. It is clear that the relative differences of GH shifts, $\Delta X_1$,

$\Delta X_5$ and $\Delta X_{10}$, are almost independent of the value of TCs, and they are also much smaller than the relative differences of IF shifts, $\Delta Y_1$, $\Delta Y_5$ and $\Delta Y_{10}$, which are increased greatly as the value of TCs increases. Note that here the polarization of light is in the TE situation.

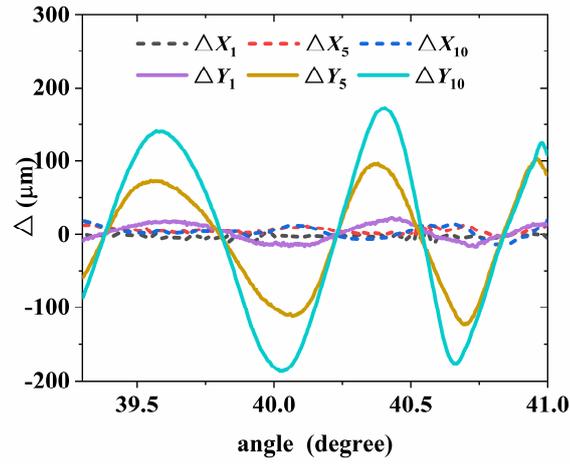

**Fig. S8.** The experimental results of the relative differences between GH shifts $\Delta X_1$, $\Delta X_5$ and $\Delta X_{10}$ and the relative differences between IF shifts $\Delta Y_1$, $\Delta Y_5$ and $\Delta Y_{10}$, for vortex beams with opposite TCs $l=\pm1, \pm5, \pm10$. Here $d = 4.55$ μm.

## References

1. M. Merano, N. Hermosa, and J. P. Woerdman, "How orbital angular momentum affects beam shifts in optical reflection," Phys. Rev. A **82**, 023817 (2010).